\documentclass[showpacs,preprintnumbers,amsmath,amssymb]{revtex4}

\usepackage{graphicx}
\usepackage{dcolumn}
\usepackage{bm}

\begin{document}

\title{Effective Action of QED in Electric Field Backgrounds II:
Spatially Localized Fields}

\author{Sang Pyo Kim}\email{sangkim@kunsan.ac.kr}
\affiliation{Department of Physics, Kunsan National University,
Kunsan 573-701, Korea} \affiliation{Asia
Pacific Center for Theoretical Physics, Pohang 790-784, Korea}

\author{Hyun Kyu Lee}\email{hyunkyu@hanyang.ac.kr}
\author{Yongsung Yoon}\email{cem@hanyang.ac.kr}
\affiliation{Physics Department, Research Institute for Natural Sciences,
Hanyang University, Seoul 133-791, Korea}

\date{\today}

\begin{abstract}
We find the Bogoliubov coefficient from the tunneling boundary
condition on charged particles in a static electric field
$E_0 \,{\rm sech}^2 (z/L)$ and, using the regularization scheme in
Phys. Rev. D {\bf 78}, 105013 (2008), obtain the exact one-loop
effective action in scalar and spinor QED. It is shown that the
effective action satisfies the general relation between the vacuum
persistence and the mean number of produced pairs. We advance
an approximation method for general electric fields and show the duality between
the space-dependent and time-dependent electric fields
of the same form at the leading order of the effective actions.
\end{abstract}
\pacs{12.20.Ds, 11.15.Tk, 13.40.-f, 12.20.-m}

\maketitle

\section{Introduction}

Understanding the vacuum structure of strong field backgrounds has
been a challenging task in quantum field theory. Electromagnetic
fields and spacetime curvatures provide a typical arena for strong
field physics. The vacuum structure may be exploited by finding the
effective action in these backgrounds. In quantum electrodynamics
(QED), Sauter, Heisenberg and Euler, Weisskopf, and Schwinger
obtained the effective action in a constant electromagnetic field
several decades ago
\cite{Sauter,Heisenberg-Euler,Weisskopf,Schwinger}. However, going
beyond the constant electromagnetic field has been another long
standing problem in QED, and the effective actions have been carried
out only for certain field configurations (for a review and
references, see Ref. \cite{Dunne} and for physical applications, see
also Ref. \cite{RVX}). For instance, there has been an attempt to
compute the effective action in a pulsed electric field of the form
$E_0 \,{\rm sech}^2 (t/\tau)$ in Refs. \cite{Dunne-Hall,KLY08}.

The main purpose of this paper is to further develop the in- and
out-state formalism based on the Bogoliubov transformation in Refs.
\cite{KLY08} (hereafter referred to I) for a pulsed electric field
and \cite{Nikishov03,Kim-Lee09} for a constant electric field to be
applicable to the case of spatially localized electric fields. To
quantize a charged particle in an electric field background is not
trivial because the vacuum is unstable against pair production.
Further, the boundary condition on the solution of the Klein-Gordon
or Dirac equation distinguishes pulsed electric fields from
spatially localized electric ones. In the former case of a pulsed
electric field, the charged boson or fermion interacts for a finite
period of time, and its positive frequency solution splits both into one branch of
positive frequency solution and into another branch of negative
frequency solution after completion of the interaction. In the
second quantized field theory, the presence of negative frequency
solution means that particle-antiparticle pairs of a given mode are
created from the vacuum due to the external electric field.

In the latter case of a spatially localized electric field,
charged bosons or fermions experience a tunneling barrier from
the Coulomb gauge potential. Nikishov elaborated the Feynman method
to find the pair-production rate in the spatially localized electric
field $E_0 \,{\rm sech}^2 (z/L)$
\cite{Nikishov70,Narozhnyi-Nikishov}. In fact, the quantum field
confronts the Klein paradox from the tunneling barrier.
To resolve the paradox, one has to treat the field in
the second quantized theory and impose a
boundary condition different from that for the
pulsed electric field \cite{Hansen-Ravndal}, which is the scattering
over the barrier. The tunneling probability through the barrier
gives the probability for one-pair production \cite{Hansen-Ravndal},
and the reflection probability leads to the vacuum persistence, that is, the
probability for the vacuum-to-vacuum transition
\cite{Padmanabhan,Srinivasan-Padmanabhan,Kim-Page02,Kim-Page06}. It
is also shown that the tunneling and the reflection probabilities
have an interpretation of instantons and anti-instantons
\cite{Kim-Page02,Kim-Page06} and that the instanton action for the
tunneling barrier provides the leading contribution to the
pair-production rate \cite{Kim-Page07}. The pair-production rate is
approximately obtained in the worldline instanton method
\cite{Dunne-Schubert,DWGS} and in the WKB method \cite{KRX}.

To calculate the effective action in $E_0 \,{\rm sech}^2 (z/L)$, we first
find the Bogoliubov coefficient as the ratio of the incident
coefficient to the reflection coefficient of the flux, whose inverse
magnitude square gives the vacuum persistence. This is consistent
with the boundary condition from causality on signals (wave packets)
of a particle \cite{Kim-Page06}. We then employ the regularization
scheme of Ref. I to calculate the effective action of scalar and
spinor QED. To our knowledge, the vacuum polarization (real part) of
the renormalized effective action is the first result for this
spatially localized field configuration.

We also advance an approximation method based on the Liouville-Green transformation
for general electric fields. The leading contributions to the effective action
in general electric fields $E(z)$ and $E(t)$ are determined entirely by the
instanton actions in Ref. \cite{Kim-Page07}. In this sense
the duality of effective actions approximately holds between any $E(z)$ and $E(t)$ of the same
function form as well as the Sauter-type electric fields.

The organization of this paper is as follows. In Sec. II, we
introduce another method to directly find the out-vacuum from the
Bogoliubov transformation without relying on the evolution operator
of two-mode squeezed operator. In Sec. III, we find the Bogoliubov
coefficient for the spatially localized electric field $E_0 \,{\rm
sech}^2 (z/L)$ and, then, compute the renormalized effective action
using the regularization scheme of Ref. I. In Sec. IV, we put forth an
approximation method to find the effective actions in general electric fields
and discuss the duality between electric fields of the same form.

\section{New Derivation of Effective Action}

In this section we develop the in- and out-state formalism for the
effective action in a background electric field. The in-state and
the out-state are related through the Bogoliubov transformation
\begin{eqnarray}
a_{\omega {\bf k}_{\perp} \sigma, {\rm out}} &=& \mu_{\omega {\bf
k}_{\perp} \sigma} a_{\omega {\bf k}_{\perp} \sigma, {\rm in}} +
\nu^*_{\omega {\bf k}_{\perp} \sigma} b^{\dagger}_{\omega {\bf
k}_{\perp} \sigma, {\rm in}}, \nonumber\\
b_{\omega {\bf k}_{\perp} \sigma, {\rm out}} &=& \mu_{\omega {\bf
k}_{\perp} \sigma} b_{\omega {\bf k}_{\perp} \sigma, {\rm in}} +
\nu^*_{\omega {\bf k}_{\perp} \sigma} a^{\dagger}_{\omega {\bf
k}_{\perp} \sigma, {\rm in}}. \label{out-in}
\end{eqnarray}
Here, $a_{\omega {\bf k}_{\perp} \sigma}$ and $b_{\omega {\bf
k}_{\perp} \sigma}$ denote the particle and the antiparticle
operators with energy $\omega$, momentum ${\bf k}_{\perp}$
transverse to the direction of the electric field, and spin $\sigma$
($\sigma =0$ for scalars and $\sigma = \pm 1/2$ for spin-1/2
fermions), respectively. The coefficients satisfy the Bogoliubov
relation
\begin{eqnarray}
| \mu_{\omega {\bf k}_{\perp} \sigma} |^2 - (-1)^{2 |\sigma|} |
\nu_{\omega {\bf k}_{\perp} \sigma}|^2 = 1.
\end{eqnarray}
In Ref. I, the Bogoliubov transformation  is expressed by the
evolution operator as
\begin{eqnarray} a_{\omega {\bf k}_{\perp} \sigma, {\rm out}} &=&
U_{\omega {\bf k}_{\perp} \sigma} a_{\omega {\bf k}_{\perp} \sigma,
{\rm in}} U^{\dagger}_{\omega {\bf
k}_{\perp} \sigma}, \nonumber\\
b_{\omega {\bf k}_{\perp} \sigma, {\rm out}} &=& U_{\omega {\bf
k}_{\perp} \sigma} b_{\omega {\bf k}_{\perp} \sigma, {\rm in}}
U^{\dagger}_{\omega {\bf k}_{\perp} \sigma}, \label{sq op}
\end{eqnarray}
and the out-vacuum is then given by $\vert 0; {\rm out} \rangle = U
\vert 0; {\rm in} \rangle$.

However, one may find the out-vacuum without using the evolution
operator. Indeed, the out-vacuum defined as
\begin{eqnarray}
a_{\omega {\bf k}_{\perp} \sigma, {\rm out}} \vert 0; {\rm out}
\rangle = 0, \quad b_{\omega {\bf k}_{\perp} \sigma, {\rm out}}
\vert 0; {\rm out} \rangle = 0,
\end{eqnarray}
is given by
\begin{eqnarray}
\vert 0; {\rm out} \rangle =  \prod_{\omega {\bf k}_{\perp} \sigma}
\Biggl[ \frac{1}{\mu_{\omega {\bf k}_{\perp} \sigma}} \sum_{n_{\omega
{\bf k}_{\perp}} = 0}^{\infty} \Bigl(- \frac{\nu^*_{\omega {\bf
k}_{\perp} \sigma}}{\mu_{\omega {\bf k}_{\perp} \sigma}}
\Bigr)^{n_{\omega {\bf k}_{\perp}}} \vert n_{\omega {\bf
k}_{\perp}}, \bar{n}_{\omega {\bf k}_{\perp}}, \sigma; {\rm in}
\rangle \Biggr]
\end{eqnarray}
for scalar QED, and by
\begin{eqnarray}
\vert 0; {\rm out} \rangle = \prod_{\omega {\bf k}_{\perp} \sigma}
\bigl[ - \nu^*_{\omega {\bf k}_{\perp} \sigma} \vert 1_{\omega {\bf
k}_{\perp}}, \bar{1}_{\omega {\bf k}_{\perp}}, \sigma, {\rm in}
\rangle + \mu_{\omega {\bf k}_{\perp} \sigma} \vert 0_{\omega {\bf
k}_{\perp}}, \bar{0}_{\omega {\bf k}_{\perp}}, \sigma; {\rm in}
\rangle \bigr]
\end{eqnarray}
for spinor QED. Here, the bar denotes the antiparticle number. The
result is the same as obtained from the evolution operator of Ref.
I. Note that particles and antiparticles are always produced in
pairs.

Then, the effective action defined by the scattering amplitude as
\begin{eqnarray}
e^{i S_{\rm eff}} = e^{i \int dt d^2{\bf x}_{\perp} {\cal L}_{\rm
eff}} = \langle 0; {\rm out} \vert 0; {\rm in} \rangle
\end{eqnarray}
is
\begin{eqnarray}
{\cal L}_{\rm eff} = (-1)^{2|\sigma|} i \sum_{\omega {\bf k}_{\perp}
\sigma} \ln (\mu^*_{\omega {\bf k}_{\perp} \sigma}). \label{eff-E}
\end{eqnarray}
Thus, the vacuum persistence follows as
\begin{eqnarray}
|\langle 0; {\rm out} \vert 0; {\rm in} \rangle |^2 = e^{ -
(-1)^{2|\sigma|} V_{\perp} T \sum_{\omega {\bf k}_{\perp}  \sigma}
\ln [1 + (-1)^{2 |\sigma|} {\cal N}_{\omega {\bf k}_{\perp}
\sigma}]},
\end{eqnarray}
where $V_{\perp}$ is the area transverse to the electric field, {\it
T} the time period, and ${\cal N}_{\omega {\bf k}_{\perp} \sigma} =
|\nu_{\omega {\bf k}_{\perp} \sigma}|^2$ is the mean number of pairs
produced. The general relation holds between the imaginary part of
the effective action and the total mean number of produced pairs
\cite{KLY08,GGT,Hwang-Kim}
\begin{eqnarray}
2 {\rm Im}({\cal L}_{\rm eff}) =  (-1)^{2|\sigma|} \sum_{\omega {\bf
k}_{\perp} \sigma} \ln [1 +(-1)^{2 |\sigma|} {\cal N}_{\omega {\bf
k}_{\perp} \sigma}]. \label{gen-rel}
\end{eqnarray}
Note that the effective action (\ref{eff-E}) and the general
relation (\ref{gen-rel}) should be renomormalized as will be shown
in the next section.

\section{Effective Action for $E(z) = E_0\,{\rm sech}^2(z/L)$}

The boundary condition for a quantum field coupled to a
space-dependent gauge field differs from that coupled to a
time-dependent gauge field. In the latter case of the time-dependent
gauge field, the in-state is a quantum state before onset of the
interaction, which evolves to an out-state after completion of
interaction. However, in the former case of the space-dependent
gauge field, though the in-/out-state cannot be defined in the
remote past/future, these may be defined analogously to the case of
time-dependent gauge field. Indeed, Nikishov developed the
scattering formalism for space-dependent gauge fields, where the
incoming signal (wave packet) toward the barrier defines the
in-state while the outgoing signal defines the out-state
\cite{Nikishov70}. Here, a caveat is that the space-dependent gauge
field confronts the Klein paradox, contrary to the time-dependent
gauge field. The resolution from causality requirement is that the
vacuum persistence (probability for the vacuum-to-vacuum transition)
is given by the reflection probability and pair production is
related to the tunneling probability
\cite{Hansen-Ravndal,Padmanabhan,Srinivasan-Padmanabhan,Kim-Page02,Kim-Page06}.
Thus, the ratio of the incident coefficient to the reflection
coefficient of the flux is the Bogoliubov coefficient for the vacuum
persistence.

As a spatially localized electric field, we consider the Sauter-type
field, $E(z) = E_0\,{\rm sech}^2(z/L)$, which extends effectively
over a length scale of $L$. In the Coulomb gauge, the gauge
potential is $A_0(z) = - E_0 L \tanh (z/L)$. The Fourier component
of the Klein-Gordon equation for scalar QED and the spin-diagonal
component of the Dirac equation for spinor QED satisfy [in units
with $\hbar = c = 1$ and with metric signature $(+, -, -, -)$]
\begin{eqnarray}
\Bigl[\partial_z^2 - (m^2 +
 {\bf k}_{\perp}^2) + (\omega - qE_{0}L \tanh (z/L))^2 +
2 i \sigma q E(z) \Bigr] \phi_{\omega {\bf k}_{\perp} \sigma} (z) = 0, \label{wave-eq}
\end{eqnarray}
where $\sigma = 0$ for scalar particles and $\sigma = \pm 1/2$ for
spin-1/2 fermions. The equation has two asymptotic momenta at $z =
\pm \infty$
\begin{eqnarray}
k_{z(\pm)} = \sqrt{(\omega \mp qE_{0}L)^{2} - m^{2}-{\bf
k}_{\perp}^{2}}.
\end{eqnarray}
Pairs are produced only when $\omega + qE_0L \geq  m$ and $\omega -
qE_0L \leq - m$, under which the particle- and antiparticle-state
can be defined asymptotically and the transverse momentum can take
the maximum ${\bf k}^2_{\perp {\rm max}} = {\rm min} \{ (\omega +
qE_0 L)^2 - m^2, (\omega - qE_0 L)^2 - m^2 \}$.

Changing the variable as
\begin{eqnarray}
\zeta = - e^{-2z/L},
\end{eqnarray}
we find the solution in terms of the hypergeometric function as
\begin{eqnarray}
\phi_{\omega {\bf k}_{\perp} \sigma} (z) = \frac{\zeta^{- iL
k_{z(+)}/2}}{\sqrt{2 k_{z(+)} e^{\pi L k_{z(+)}}}} (1 - \zeta)^{(1 -
2 \sigma)/2 + i \lambda_{\sigma}} F(\alpha_{\omega {\bf
k_{\perp}}\sigma}, \beta_{\omega {\bf k_{\perp}}\sigma};
\gamma_{\omega {\bf k_{\perp}}}; \zeta), \label{sol}
\end{eqnarray}
where
\begin{eqnarray}
\lambda_{\sigma} = \sqrt{(qE_0 L^2)^2 - \Bigl(\frac{1 -
2|\sigma|}{2} \Bigr)^2},
\end{eqnarray}
and
\begin{eqnarray}
\alpha_{\omega {\bf k_{\perp}} \sigma} &=& \frac{1- 2 \sigma}{2} -
\frac{i}{2} \bigl(L k_{z(+)} - L k_{z(-)}
- 2 \lambda_{\sigma} \bigr), \nonumber \\
\beta_{\omega {\bf k_{\perp}}\sigma} &=& \frac{1- 2 \sigma}{2} -
\frac{i}{2}
\bigl(L k_{z(+)} + L k_{z(-)}  - 2 \lambda_{\sigma} \bigr), \nonumber \\
\gamma_{\omega {\bf k_{\perp}}} &=& 1 - i L k_{z(+)}.
\end{eqnarray}
The solution is normalized to have the asymptotic form at $z =
\infty$,
\begin{eqnarray}
\phi_{\omega {\bf k_{\perp}}} (z) = \frac{ e^{i k_{z(+)} z} }{
\sqrt{2 k_{z (+) } } } .
\end{eqnarray}
From the transformation formula \cite{GR},  the other asymptotic
form at $z= - \infty$ is given by
\begin{eqnarray}
\phi_{\omega {\bf k_{\perp}}\sigma}(z) = A_{\omega {\bf k_{\perp}}
\sigma} \frac{ e^{i k_{z(-)} z} }{\sqrt{2 k_{z(+)}}} + B_{\omega
{\bf k_{\perp}} \sigma} \frac{ e^{-i k_{z(-)} z} }{ \sqrt{2 k_{z(+)}
}},
\end{eqnarray}
where the incident and the reflection coefficients are
\begin{eqnarray}
A_{\omega {\bf k_{\perp}}\sigma} &=& \frac{ \Gamma (\gamma_{\omega
{\bf k_{\perp}}}) \Gamma(\beta_{\omega {\bf k_{\perp}}\sigma} -
\alpha_{\omega {\bf k_{\perp}}\sigma})}{ \Gamma ( \beta_{\omega {\bf
k_{\perp}}\sigma}) \Gamma (\gamma_{\omega {\bf k_{\perp}}} -
\alpha_{\omega {\bf k_{\perp}}\sigma})}, \nonumber
\\ B_{\omega {\bf k_{\perp}} \sigma} &=&
\frac{ \Gamma (\gamma_{\omega {\bf k_{\perp}}})
\Gamma(\alpha_{\omega {\bf k_{\perp}}\sigma} - \beta_{\omega {\bf
k_{\perp}}\sigma})}{ \Gamma ( \alpha_{\omega {\bf k_{\perp}}\sigma})
\Gamma (\gamma_{\omega {\bf k_{\perp}}} - \beta_{\omega {\bf
k_{\perp}}\sigma})}.
\end{eqnarray}
The Bogoliubov coefficient, $\mu_{\omega {\bf k_{\perp}}\sigma} =
A_{\omega {\bf k_{\perp}}\sigma}/B_{\omega {\bf k_{\perp}}\sigma}$,
from the group velocity can be written as
\begin{eqnarray}
\mu^*_{\omega {\bf k_{\perp}}\sigma} = \frac{\Gamma (\frac{1-2
\sigma}{2} + i \frac{ \Delta^{(-)} }{2})}{\Gamma (\frac{1-2
\sigma}{2} + i \frac{\Omega^{(-)}}{2} )}  \times \frac{\Gamma
(\frac{1+ 2 \sigma}{2} + i \frac{\Delta^{(+)}}{2} )}{\Gamma
(\frac{1+ 2 \sigma}{2} + i \frac{\Omega^{(+)}}{2} )}, \label{mu-sp}
\end{eqnarray}
where
\begin{eqnarray}
\Omega_{\omega{\bf k_{\perp}} \sigma}^{(\pm)} &=& L k_{z(+)} + L
k_{z(-)} \pm 2 \lambda_{\sigma}, \nonumber\\
\Delta_{\omega{\bf k_{\perp}}\sigma}^{(\pm)} &=& L k_{z(+)} - L
k_{z(-)} \pm 2 \lambda_{\sigma}.
\end{eqnarray}
Here, we have deleted the term, $\Gamma (-i L k_{z(-)})/\Gamma (iL
k_{z(-)})$, which is independent of the interaction with the
electric field and is removed through normalization. Note that
 $\Omega_{\omega {\bf k_{\perp}}}^{(+)}, \Delta_{\omega
{\bf k_{\perp}}}^{(+)} > 0$ and $\Omega_{\omega {\bf
k_{\perp}}}^{(-)}, \Delta_{\omega {\bf k_{\perp}}}^{(-)} < 0$.

Now, we compute the effective action in Eq. (\ref{eff-E}). For that
purpose, we follow the method of Ref. I, where we use the integral
representation of the gamma function \cite{gamma-func}, sum over two
spin states, $\sigma = \pm 1/2$, do the contour integral of the
first term in the first quadrant and that of the second term in the
fourth quadrant, and subtract the divergent terms,
which is equivalent to renormalizing the vacuum energy
and the charge. Finally, we obtain the exact one-loop
effective action of scalar QED per unit time and per unit
cross-sectional area
\begin{eqnarray}
{\cal L}^{\rm sc}_{\rm eff} &=& \frac{1}{2} \int \frac{d\omega
d^2{\bf k}_{\perp}}{(2 \pi)^3} {\cal P} \int_0^{\infty} \frac{ds}{s}
(e^{\Delta^{(-)}_{\omega {\bf k_{\perp}}} s} - e^{-
\Delta^{(+)}_{\omega {\bf k_{\perp}}} s} +
 e^{- \Omega^{(+)}_{\omega {\bf k_{\perp}}} s}
 - e^{\Omega^{(-)}_{\omega
{\bf k_{\perp}}} s} ) \Bigl( \frac{1}{\sin(s)} - \frac{1}{s} -
\frac{s}{6} \Bigr) \nonumber\\ && + \frac{i}{2} \int \frac{d\omega
d^2{\bf k}_{\perp}}{(2 \pi)^3} \ln \Biggl[ \frac{ \cosh (\pi
\Omega^{(+)}_{\omega{\bf k_{\perp}}}/2) \cosh (\pi
\Omega^{(-)}_{\omega{\bf k_{\perp}}}/2)}{\cosh (\pi
\Delta^{(+)}_{\omega{\bf k_{\perp}}}/2)\cosh (\pi
\Delta^{(-)}_{\omega{\bf k_{\perp}}}/2)} \Biggr], \label{sc-eff}
\end{eqnarray}
and that of spinor QED
\begin{eqnarray}
{\cal L}^{\rm sp}_{\rm eff} &=& - \int \frac{d\omega d^2{\bf
k}_{\perp}}{(2 \pi)^3} {\cal P} \int_0^{\infty} \frac{ds}{s}
(e^{\Delta^{(-)}_{\omega {\bf k_{\perp}}} s} - e^{-
\Delta^{(+)}_{\omega {\bf k_{\perp}}} s} +
 e^{- \Omega^{(+)}_{\omega {\bf k_{\perp}}} s}
 - e^{\Omega^{(-)}_{\omega
{\bf k_{\perp}}} s}) \Bigl( \cot (s) - \frac{1}{s} + \frac{s}{3} \Bigr) \nonumber\\
&& - i \int \frac{d\omega d^2{\bf k}_{\perp}}{(2 \pi)^3} \ln \Biggl[
\frac{ \sinh (\pi \Omega^{(+)}_{\omega{\bf k_{\perp}}}/2) \sinh (\pi
\Omega^{(-)}_{\omega{\bf k_{\perp}}}/2)}{\sinh (\pi
\Delta^{(+)}_{\omega{\bf k_{\perp}}}/2)\sinh (\pi
\Delta^{(-)}_{\omega{\bf k_{\perp}}}/2)} \Biggr]. \label{sp-eff}
\end{eqnarray}
Here, the integration is restricted to $\int d \omega =
\int_{-(qE_0L -m)}^{qE_0L-m} d\omega$ and $\int d^2{\bf k}_{\perp} =
2 \pi \int_0^{k_{\perp {\rm max}}} k_{\perp {\rm max}} d k_{\perp
{\rm max}}$. It can be shown that the general relation between the
vacuum persistence (twice of the imaginary part) and the mean number
of produced pairs holds in scalar and spinor QED
\begin{eqnarray}
2 {\rm Im} ({\cal L}_{\rm eff}) = (-1)^{2 |\sigma|} \int
\frac{d\omega d^2{\bf k}_{\perp}}{(2 \pi)^3} \ln (1 + (-1)^{2
|\sigma|} {\cal N}_{\omega{\bf k_{\perp}} \sigma}), \label{vac-mean}
\end{eqnarray}
where
\begin{eqnarray}
{\cal N}_{\omega{\bf k_{\perp}} \sigma} = \frac{2 \sinh (\pi L
k_{z(+)}) \sinh (\pi L k_{z(-)})}{\cosh (2 \pi \lambda_{\sigma}) +
(-1)^{2 \sigma} \cosh(\pi L k_{z(+)}- \pi L k_{z(-)})}.
\label{mean-num}
\end{eqnarray}

A few comments are in order. First, the mean number of produced pairs, Eq.
(\ref{mean-num}), agrees with the exact result of Refs.
\cite{Nikishov70,Hansen-Ravndal} and also Ref.
\cite{Chervyakov-Kleinert} for scalar QED. The effective action in a
constant electric field can be obtained by taking $L = \infty$. In fact, the term
$e^{\Omega^{(-)}_{\omega {\bf k_{\perp}}} s}$ yields the constant
field limit while all the other terms vanish. The leading term of Eq.
(\ref{mean-num}), ${\cal N}_{\omega {\bf k_{\perp}} \sigma} \approx
e^{- \pi (2 \lambda_{\sigma} - L k_{z(+)} - L k_{z(+)})}$, agrees
with Eq. (36) of Ref. \cite{Kim-Page07} from the instanton action.
Second, it would be interesting to compare the effective actions (\ref{sc-eff}) and
(\ref{sp-eff}) with Eqs. (66) and (80) for $E(t) =
E_0\,{\rm sech} (t/\tau)$ in Ref. I. The kinetic momenta $k_{z(\pm)}$ along the direction of
the electric field at spatial infinities
now correspond to the kinetic energy $\omega_{{\bf k} (\pm)}$ at the remote past and future.
However, the different boundary conditions select different contours so
that the mean number of produced pairs, Eq. (\ref{mean-num}), becomes
the inverse of Eqs. (68) and (83) of Ref. I. This point will be further discussed
 in the next section.

\section{Approximate Effective Actions in General Electric Fields}

We now put forth an approximation method for the effective action
in general electric fields, which cannot be solved exactly.
In a general Coulomb potential $A_0(z)$, the Fourier component
of the field equation (\ref{wave-eq}) is given by
\begin{eqnarray}
\Biggl[\frac{d^2}{dz^2} + Q_{\omega {\bf k}_{\perp} \sigma} (z) \Biggr]
\phi_{\omega {\bf k}_{\perp} \sigma} (z) = 0, \label{gen wave}
\end{eqnarray}
where
\begin{eqnarray}
Q_{\omega {\bf k}_{\perp} \sigma} (z) =
(\omega + qA_0(z))^2 - (m^2 +
 {\bf k}_{\perp}^2)  +
2 i \sigma q E(z).
\end{eqnarray}
Here $|\omega + qA_0(\pm \infty)| \geq m$ and the maximum transverse
momentum ${\bf k}^2_{\perp {\rm max}} = {\rm min} \{ (\omega + qA_0(\pm \infty))^2 - m^2 \}$.
Our stratagem is to transform Eq. (\ref{gen wave}) into the differential equation
whose solution can be found approximately. In Ref. \cite{Dunne-Hall}
the uniform semiclassical approximation is used for general time-dependent electric fields.
The uniform semiclassical approximation \cite{BFH} is an extension of the
Liouville-Green transformation \cite{Kim92} to the following form
\begin{eqnarray}
\Biggl[ \frac{d^2}{d \eta^2} + \eta^2 - \frac{{\cal S}_{\omega {\bf k}_{\perp} \sigma}}{\pi}
 + \frac{1}{(d \eta/dz)^{3/2}} \frac{d^2}{dz^2} \Bigl(\frac{1}{\sqrt{d \eta/dz}} \Bigr)
  \Biggr] \varphi_{\omega {\bf k}_{\perp} \sigma} (\eta) = 0, \label{LG}
\end{eqnarray}
where $\varphi_{\omega {\bf k}_{\perp} \sigma} (\eta) = \sqrt{d \eta/dz} \phi_{\omega {\bf k}_{\perp} \sigma} (z)$ and
\begin{eqnarray}
\Bigl(\eta^2 - \frac{{\cal S}_{\omega {\bf k}_{\perp} \sigma}}{\pi}  \Bigr) \Bigl( \frac{d \eta}{dz} \Bigr)^2 = Q_{\omega {\bf k}_{\perp} \sigma} (z). \label{eta con}
\end{eqnarray}
Doing a contour integral exterior to the branch cut \cite{Markushevich},  ${\cal S}_{\omega {\bf k}_{\perp} \sigma}$
turns out to be the instanton action in $E(z)$ \cite{Kim-Page07}:
\begin{eqnarray}
{\cal S}_{\omega {\bf k}_{\perp} \sigma} = - i \oint dz \sqrt{Q_{\omega {\bf k}_{\perp} \sigma}(z)}.
\label{ins ac1}
\end{eqnarray}

The electric field studied in this paper is either constant or spatially localized such that
$A_0 (z) \propto z^{1-c}$ with $c \geq 0$ for $|z| \gg 1$ and less singular than $1/z$ at finite $z$.
Then the correction term, the last term in Eq. (\ref{LG}), is asymptotically proportional
to $1/ \eta^2$ and may be neglected in this approximation scheme. Thus,
we approximately find the transmitted wave in terms of the parabolic cylinder function
\begin{eqnarray}
\varphi_{\omega {\bf k}_{\perp} \sigma} (\eta) = D_p (\sqrt{2} e^{i \pi/4} \eta ), \quad p = - \frac{1}{2} + \frac{i}{2\pi} {\cal S}_{\omega {\bf k}_{\perp} \sigma}, \label{par cyl}
\end{eqnarray}
which has the asymptotic form $D_p (\sqrt{2} e^{i \pi/4} \eta ) \propto \eta^p e^{-i \eta^2 /2}$ for $\eta \gg 1$. On the other hand, for $\eta \ll -1$ the solution (\ref{par cyl})
has another asymptotic form
\begin{eqnarray}
\varphi_{\omega {\bf k}_{\perp} \sigma} (\eta) = \frac{\sqrt{2 \pi} e^{-i (p+1) \pi /2}}{\Gamma(-p)} D_{-(p+1)} (\sqrt{2} e^{i 3\pi/4} \eta ) + e^{-i p \pi} D_p (\sqrt{2} e^{i 5\pi/4} \eta ),
\end{eqnarray}
and the ratio of the incident coefficient to the reflection coefficient for the flux is the Bogoliubov coefficient
\begin{eqnarray}
\mu_{\omega {\bf k}_{\perp} \sigma} = \frac{\sqrt{2 \pi}}{\Gamma (-p)} e^{i(p-1) \pi/2}.
\end{eqnarray}
Following the procedure in Sec. III and doing the contour integral in the fourth quadrant,
we obtain the approximate effective action
\begin{eqnarray}
{\cal L}_{\rm eff} = \frac{(-1)^{2 |\sigma|} }{2}  \sum_{\sigma} \int \frac{d\omega d^2{\bf
k}_{\perp}}{(2 \pi)^3} \Biggl[  {\cal P} \int_0^{\infty} \frac{ds}{s}
e^{- s {\cal S}_{\omega {\bf k}_{\perp} \sigma}/\pi} \Bigl(\frac{1}{\sin (s)} - \cdots \Bigr)
- i \sum_{n = 1}^{\infty} \frac{(-1)^{n}}{n} e^{- n {\cal S}_{\omega {\bf k}_{\perp} \sigma}}
\Biggr]. \label{app eff}
\end{eqnarray}
Here dots denote the terms to regularize the vacuum energy and the charge.
For a constant electric field, $A_0 (z) = - E_0 z$ and ${\cal S}_{\omega {\bf k}_{\perp} \sigma} = \pi (m^2 +
 {\bf k}_{\perp}^2 - 2 i \sigma q E_0)/qE_0$, so the effective action (\ref{app eff}) recovers the
 Heisenberg-Euler effective action. For $E(z) = E_0\,{\rm sech}^2(z/L)$, ${\cal S}_{\omega {\bf k}_{\perp} \sigma} \approx - \pi \Omega^{(-)}_{\omega {\bf k}_{\perp} \sigma =1/2} + 2 \pi i \sigma$ with $\Omega^{(-)}_{\omega {\bf k}_{\perp} \sigma =1/2} < 0$, and Eq. (\ref{app eff}) becomes for scalar QED
\begin{eqnarray}
{\cal L}^{\rm sc} _{\rm eff} = \frac{1}{2} \int \frac{d\omega d^2{\bf
k}_{\perp}}{(2 \pi)^3} \Biggl[  {\cal P} \int_0^{\infty} \frac{ds}{s}
e^{s \Omega^{(-)}_{\omega {\bf k}_{\perp} \sigma =1/2}} \Bigl(\frac{1}{\sin (s)} - \frac{1}{s} -
\frac{s}{6} \Bigr)
+ i \ln (1 + e^{\pi \Omega^{(-)}_{\omega {\bf k}_{\perp}\sigma =1/2}})
\Biggr], \label{app sc eff}
\end{eqnarray}
and for spinor QED
\begin{eqnarray}
{\cal L}^{\rm sp} _{\rm eff} = - \int \frac{d\omega d^2{\bf
k}_{\perp}}{(2 \pi)^3} \Biggl[  {\cal P} \int_0^{\infty} \frac{ds}{s}
e^{s \Omega^{(-)}_{\omega {\bf k}_{\perp}\sigma =1/2}} \Bigl(\cot(s) - \frac{1}{s} +
\frac{s}{3} \Bigr)
+ i \ln (1 - e^{\pi \Omega^{(-)}_{\omega {\bf k}_{\perp} \sigma =1/2} })
\Biggr]. \label{app sp eff}
\end{eqnarray}
The results, (\ref{app sc eff}) and (\ref{app sp eff}), are consistent with
the leading terms of Eqs. (\ref{sc-eff}) and (\ref{sp-eff}), respectively.

The approximation method can also be applied to the time-dependent electric fields $E(t)$.
The Fourier component of the field equation in the gauge field $A_z (t)$ takes the form
\begin{eqnarray}
\Biggl[\frac{d^2}{dt^2} + Q_{{\bf k} \sigma} (t) \Biggr]
\phi_{{\bf k} \sigma} (t) = 0,
\end{eqnarray}
where
\begin{eqnarray}
Q_{{\bf k} \sigma} (t) =
(k_z + qA_z(t))^2 + (m^2 +
 {\bf k}_{\perp}^2)  +
2 i \sigma q E(t).
\end{eqnarray}
Changing the variable
\begin{eqnarray}
(\xi^2 + \frac{{\cal S}_{{\bf k} \sigma}}{\pi} ) \Bigl( \frac{d \xi}{dt} \Bigr)^2 = Q_{{\bf k} \sigma} (t),
\end{eqnarray}
and introducing the instanton action \cite{Kim-Page07}
\begin{eqnarray}
{\cal S}_{{\bf k} \sigma} =  i \oint dt \sqrt{Q_{{\bf k} \sigma}(t)}, \label{ins ac2}
\end{eqnarray}
and finally doing the contour integral in the first quadrant, we approximately obtain the effective actions in Sec. III of Ref. I. There the only modification is the parameter $p = -1/2 - i{\cal S}_{{\bf k} \sigma}/2\pi$.
Thus, the leading contribution to the effective action is the same as Eq. ({\ref{app eff}) with ${\cal S}_{{\bf k} \sigma}$ replacing ${\cal S}_{\omega {\bf k}_{\perp} \sigma}$.

A passing remark is that the approximation method based on the Liouville-Green transformation
not only provides the effective action (\ref{app eff}) but also explains how the instanton actions
(\ref{ins ac1}) and (\ref{ins ac2}) determine the mean number of the produced pairs
either in spatially localized electric fields or in pulsed electric fields, as shown in Ref. \cite{Kim-Page07}. The different boundary conditions for electric fields
$E(z)$ and $E(t)$, which are imprinted in the parameters $p=-1/2 \pm i{\cal S}_{{\bf k} \sigma}/2\pi$, requires contours in the fourth and first quadrant, respectively.
As a consequence, the duality approximately holds between $E(z)$ and $E(t)$ for the same form.
Further, the mean numbers of produced pairs for $E(z) = E_0\,{\rm sech}^2(z/L)$ and
$E(t) = E_0\,{\rm sech}^2(t/\tau)$ are inverse to each other in the form.

\section{Conclusion}

In this paper we have further developed the regularization scheme
using the Bogoliubov coefficient in Ref.
I to obtain the effective action in $E(z) = E_0\,{\rm sech}^2(z/L)$.
The Klein paradox due to tunneling barrier makes the
boundary condition on the field equation in a space-dependent gauge
differ from that in a time-dependent gauge. This is resolved by the causality
argument, which requires the reflection and the transmission
probabilities not by the flux but by the group velocity or signal. The
Bogoliubov coefficient is then given by the ratio of the incident
coefficient to the reflection one with respect to the flux, which
is the reason why the mean number of produced pairs in $E(z) = E_0\,{\rm sech}^2(z/L)$ is inverse of
that in $E(t) = E_0\,{\rm sech}^2(t/\tau)$.

We have also introduced an approximation method for general electric fields, which
is based on the Liouville-Green transformation that changes the field equation into a solvable one,
for instance, the parabolic equation with correction terms. The method can be applied both to
spatially localized electric fields $E(z)$ and to pulsed electric fields $E(t)$.
Remarkably, the leading contributions are determined by the instanton actions, which confirm
the mean number of produced pairs in  the phase-integral approximation \cite{Kim-Page07}.
Further, the leading contribution to the effective actions show duality between the space-dependent and
time-dependent Sauter type electric fields and the duality seems to be generic for $E(z)$
and $E(t)$ of the same form at this approximation.
However, whether this exists the exact duality remains an open question.

\acknowledgments

S.~P.~K.~would like to thank Gerald V.~Dunne for useful comments on the
duality between the time-dependent and space-dependent Sauter-type
electric fields, and W-Y.~Pauchy Hwang, Wei-Tou Ni, and Don N.~Page for
critically reading the manuscript. He also appreciates the warm
hospitality of the Center for Quantum Spacetime (CQUeST) of Sogang
University, Purple Mountain Observatory, Chinese Academy of
Sciences, the Leung Center for Cosmology and Particle
Astrophysics (LeCosPA) of National Taiwan University, and
Yukawa Institute for Theoretical Physics of Kyoto University. This work was
supported by the Korea Research Foundation (KRF) Grant funded by the
Korean Government (MEST)(2009-0075-773).
\appendix

\end{document}